\documentclass[sigconf]{acmart}
\usepackage{amsmath,amssymb,amsfonts}
\usepackage{moreverb,algorithm,algorithmic}
\usepackage{graphicx}
\usepackage{caption}

\setcopyright{acmcopyright}
\copyrightyear{2019}
\acmYear{2019}
\acmDOI{10.1145/3366486.3366523}
\acmConference[To Appear in WUWNET'19]{WUWNET'19: International Conference on Underwater Networks and Systems}{October 23--25, 2019}{Atlanta, GA, USA}

\begin{document}

\title{Real-time Image Enhancement for Vision-based Autonomous Underwater Vehicle Navigation in Murky Waters}
\author{
Wenjie Chen, Mehdi Rahmati, Vidyasagar Sadhu, and Dario Pompili}
\affiliation{Department of Electrical and Computer Engineering, Rutgers University--New Brunswick, NJ, USA}
\email{{wenjie.chen, mehdi.rahmati, vidyasagar.sadhu,  pompili}@rutgers.edu}

\pagestyle{plain}
\pagenumbering{arabic}
\thispagestyle{empty}
\pagestyle{empty}

\begin{abstract}
Classic vision-based navigation solutions,
which are utilized in algorithms such as Simultaneous Localization and Mapping~(SLAM), usually fail to work underwater when the water is murky and the quality of the recorded images is low. That is because most SLAM algorithms are feature-based techniques and often it is impossible to extract the matched features from blurry underwater images. 
To get more useful features, image processing techniques can be used to dehaze the images before they are used in a navigation/localization algorithm. There are many well-developed methods for image restoration, but the degree of enhancement and the resource cost of the methods are different. In this paper, we propose a new visual SLAM, specifically-designed for the underwater environment, using Generative Adversarial Networks~(GANs) to enhance the quality of underwater images with underwater image quality evaluation metrics. This procedure increases the efficiency of SLAM and gets a better navigation and localization accuracy. We evaluate the proposed GANs-SLAM combination by using different images with various levels of turbidity in the water. Experiments were conducted and the data was extracted from the Carnegie Lake in Princeton, and the Raritan river both in New Jersey, USA. 
\end{abstract}

\maketitle

\section{Introduction}

\textbf{Overview:}
In recent years, a wide range of underwater applications---such as multimedia coastal and tactical surveillance, offshore exploration, seafloor mapping, submarine volcanism, and hydro-thermal vent studies---have been developed that require multimedia data to be captured, retrieved, and processed reliably in real time~\cite{rahmati2017ssfb}. In many of these applications, Unmanned or Autonomous Underwater Vehicles~(UUVs, AUVs), equipped with the camera and multiple on-board sensors, are used in the exploration of undersea resources and for gathering scientific data in autonomous or semi-autonomous monitoring missions~\cite{rahmati2019uwsvc}.

Camera-equipped underwater vehicles, which are able to use computer vision algorithms~\cite{rahmati2018adaptive}, are an important group of underwater vehicles, since Global Position System~(GPS) signal is not available underwater and hence some of AUVs try to navigate by using the camera via visual positioning techniques such as Simultaneous Localization and Mapping~(SLAM)~\cite{rahmatislam2018}. In general, the navigation for AUVs can be divided into three popular types: inertial navigation, acoustic transponders and modems, and Geophysical navigation. The inertial navigation uses the information from the accelerometers and gyroscopes to estimate the current state based on algorithms such as Dead Reckoning~(DR). The main problem of this method is that the position error is cumulative and grows unbounded. Therefore, there will be a huge deviation between the real position and the estimated results after a few minutes. The method uses acoustic transponders and modems to measures the travel-time of signals that are transmitted from acoustic beacons to estimate the current position. 
These methods take advantage of additional objects with exact locations to estimate the movement of AUVs. Geophysical navigation uses sensors such as sonars and cameras to get the external environmental information. This type of navigation should be capable of detecting, identifying and classifying environmental features~\cite{Sadhu2019tmc}. GPS signal will be lost underwater even though certain expensive GPS equipment can work with a limited range~\cite{Sadhu2019percom}. 

Moreover, we want the algorithm to be used for unknown environments~\cite{Sadhu2016icac, sadhu2019iros} so that there is no need to use pre-deployed transponders and modems. In such conditions, a vision-based algorithm is a good solution to explore the environments including seabed and the bottom of the rivers, in which SLAM is used for navigation and localization purposes. 
Early research on SLAM uses laser or ultrasonic sensors as inputs. With the development of computer vision, the visual SLAM has gained attraction for robot localization and mapping needs. Compared with laser or ultrasonic sensors, robots can obtain more useful information from low-cost video cameras. Powerful feature detectors and descriptors like ORB~\cite{rublee2011orb} can be used for SLAM. 


\textbf{Motivation:}
Generally, the water in the rivers is muddy, especially after rain or flood, as the particles from the surrounding land are washed into the river. This makes the water to have a higher turbidity value. If we directly use the images from the camera for the navigation, the system will fail. Moreover, underwater images usually face the problem of the distortion due to the light scattering and absorption, which leads to the lack of contrast, blurry appearance, and inaccurate color detection. The red and orange wavelengths of lights are absorbed by the water more quickly than the blue and green light, hence the image is often in a bluish-green tone. While the absorption and the scattering are not the only factors that affect the quality of the images, water pollution and floating particles also will increase the complexity of the image recovery. The main challenge---before using underwater images in SLAM---is to remove the turbidity and restore the color as per the underwater environment. Therefore, the processing of underwater images is divided into two main categories: underwater image enhancement and underwater image restoration. Image enhancement methods focus on improving the image features and RGB values to improve the quality of the image, while image restoration aims at reducing the distortion effect caused by the underwater environments. There are some early works that rely on using additional information such as polarization-based method~\cite{liang2014visibility}, which uses different degrees of polarization to remove the haze. However, requiring many physical parameters makes it inflexible and hence it is not recommended. A method called Dark Channel Prior~(DCP), proposed by He~\cite{he2011single}, has a great improvement for haze removal in the air. On the account of the similarities of image processing, DCP has been widely used in the field of underwater images restoration.
Based on the good results obtained by DCP, a new method is proposed called UDCP~\cite{drews2016underwater} that only considers the blue and green channels instead of all channels in DCP.


Because of the success of deep learning methods in computer vision, there are some research that try to use deep learning in underwater image processing. Generative Adversarial Networks~(GANs) have been a topic of recent interest in deep learning that are shown to be good at the image to image translation. To solve the problems related to underwater image degradation, WaterGAN~\cite{li2018watergan} is proposed that creates realistic underwater images from the in-air image and depth pairings in an unsupervised pipeline for color correction of monocular underwater images. This method requires a large paired image datasets. Moreover, there is another GAN-based work, Multi-scale Adversarial Network~\cite{lu2019multi}, which is based on CycleGAN~\cite{zhu2017unpaired} with adaptive size window for Structural SIMilarity~(SSIM) loss.

\textbf{Our Approach:}
Due to the limited energy of AUVs, there is a need to find the trade-off between the energy cost and the performance for different methods. Moreover, the degree of the image enhancement using different methods is not the same. Therefore, different methods can be used to improve underwater images with different clarity levels. We propose a new method based on image quality evaluation metrics that combines the advantages of different methods and overcomes their drawbacks. First, we present different methods to recover images and compare their performance. Second, we find a proper parameter that will be used to determine the level of enhancement of images. Finally, we find the critical point of parameter values and use the proper method to improve different images with different clarity levels.

\textbf{Our Contribution:}
There are many datasets for the marine environment; but a few datasets for rivers/lakes are available. Hence, in this research, first, we create an underwater image dataset for the Carnegie Lake, Princeton, New Jersey and the Raritan River, Somerset, New Jersey. This dataset contains multiple varieties of under river/lakes situations that includes both blurry and corresponding clear images, which can be used for future underwater research. Second, we propose a new method to enhance underwater images especially blurry images (tainted by the murky water). Using underwater image quality evaluation metrics, 
we come up with a critical value for each image and get better results with an acceptable efficiency and a low power consumption. Using our enhanced images, we improve the accuracy of the SLAM algorithm.  

\textbf{Paper Outline:}
In Sect.~\ref{sec:related}, we discuss the related work and position our solution w.r.t. the literature. In Sect.~\ref{sec:method}, we introduce our haze-removal real-time strategies aimed at enhancing images for vision-based autonomous underwater vehicle navigation in murky waters. Then, in Sect.~\ref{sec:expr}, we discuss the simulation and experiment setup, and present the performance evaluation results. Finally, in Sect.~\ref{sec:conc}, we draw the main conclusions and discuss future work.

\section{Related Work}\label{sec:related}
In this section, we present first various existing underwater image enhancement methods (Sect.~\ref{sec:relwork:enhance}); then, we provide some background on different underwater image-quality metrics that are used to evaluate image-enhancement solutions (Sect.~\ref{sec:relwork:evaluate}). Finally, we discuss state-of-the-art SLAM-based algorithms (Sect.~\ref{sec:relwork:slam}).


\subsection{Underwater Image Enhancement Methods}\label{sec:relwork:enhance}
Exploring underwater environment has seen great development recently, and the use of underwater image enhancement solutions has been a key technique towards this goal.
There are many methods
in the literature that can be divided into the following groups.

\textit{Supplementary information-based methods:}
In these methods, the supplementary information from multiple images is used to improve the quality of underwater images including polarization filtering~\cite{schechner2005recovery} and range-gated imaging~\cite{li2009speckle}. It is necessary to take multiple images to get sufficient information, which makes it more complicated than a single image method.

\textit{Non-physical model-based methods:}
This method adjusts the image pixel values to improve the quality of images. Authors in~\cite{iqbal2010enhancing}  improve the contrast and the saturation of a single image by changing the pixel range in Red Green Blue~(RGB) and Hue Saturation Value~(HSV) color spaces.

\textit{Physical model-based methods:}
This method collects the parameters of an image model from an input image and applies the model to the image to improve the quality. One direction is based on the Dark Channel Prior~(DCP)~\cite{he2011single}. Chiang et al.~\cite{chiang2012underwater} propose a method that combines the DCP and the wavelength-dependent compensation to restore images. Drews et al.~\cite{drews2016underwater} propose Underwater DCP by only ignoring the information of red channel. Peng et al.~\cite{peng2018generalization} propose a Generalized DCP by incorporating adaptive color correction into an image formation model. Another direction is to use optical properties of underwater images. Berman et al.~\cite{berman2017diving} estimate two additional global parameters to reduce the problem of underwater image restoration to single image dehazing. Zhao et al.~\cite{zhao2015deriving} propose a method that uses inherent optical properties of water from background color to enhance the degraded underwater images.

\textit{Data-driven methods:}
This method usually utilizes a Convolutional Neural Network~(CNN) to train the model by collecting a huge amount of data. The main motivation behind such huge data is to obtain the ground truth images. Therefore, the WaterGAN, proposed in~\cite{li2018watergan}, can simulate the realistic underwater images from in-air images and depth pairing in an unsupervised pipeline. However, for this method to be effective, the dataset should contain and represent information pertaining to variety of situations. 


\subsection{Underwater Image Quality Evaluation}\label{sec:relwork:evaluate}
To compare different image enhancement solutions, the image quality evaluation metric will be used in this paper. Specifically, we discuss two categories of metrics, i.e., full-reference and non-reference. 

\textit{Full-reference Metrics:}
This method calculates natural visual characteristics --- such as Mean Square Error~(MSE), Peak Signal-to-Noise Ratio~(PSNR), and SSIM---from images and compares the value with the ground truth. Although a disk with different colors can be chosen as ground truth, it is not easy to get it done for underwater images. Moreover, sometimes, the result will be different from the human visual observation.

\textit{Non-reference Metrics:}
Generally, the ground truth cannot be obtained in a real world underwater environment. In such cases, the non-reference image quality evaluation metrics, such as image contrast, can be used. For different applications, researchers use image sharpness, slope of edges, the gradient magnitude histogram from images, and image entropy to directly measure the image quality without knowing the type of image distortion. Moreover, the result of image enhancement will finally be evaluated by human observation, i.e., Human Visual System~(HVS). The HVS is sensitive to the difference of color, edge structure, and relative contrast so that colorfulness, sharpness, and contrast can be used as the basic attributes based on human observations to propose new metrics such as Underwater Image Quality Measure~(UIQM) and Underwater Image Sharpness Measure~(UISM). Sometimes, image quality should be measured by all parameters, as  in~\cite{panetta2016human}, in which the authors propose a new UIQM metric
to find a linear combination of colorfulness, sharpness, and contrast.

\subsection{SLAM Algorithms}\label{sec:relwork:slam}
There are several types of visual SLAM methods in use currently.
The common methods for implementing the SLAM are filters, graphs, and bundle adjustment methods. In filter-based methods, Kalman filter~\cite{bailey2006consistency} based techniques are widely used, which have good performance for estimating the state of nonlinear systems and integrating different  outputs of several sensors. Particle filter is another useful method which can deal with a large number of features, but its run time increases as the map size increases. Graph-based SLAM builds the graph and finds a node configuration that minimizes the error introduced by the constraints. The main disadvantage of this method is that it consumes large amount of memory for computations. The ORB-SLAM~\cite{mur2017orb} is one of the impressive feature-based techniques, which is a complete SLAM system for monocular, stereo and RGB-D cameras. The test results show ORB-SLAM to be the most accurate SLAM solution compared with other visual SLAM systems. For ORB-SLAM, one of the main design requirements is its feature design. It uses ORB features, which are oriented multi-scale FAST corners with an associated 256 bits descriptor. Though it is fast to match, ORB-SLAM still cannot work efficiently as it is difficult to find the matching features.

\section{Our Proposed Solution}\label{sec:method}
\begin{figure}[!t]
\centering
\includegraphics[scale=0.65]{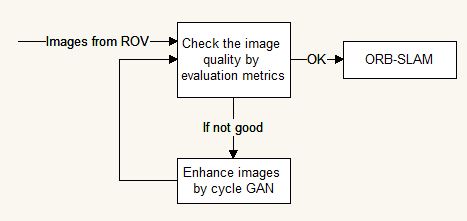}\\
\caption{\label{fig2}High-level block diagram of our solution. 
}
\end{figure}

The physical model-based methods and the data-driven methods are used widely for underwater image enhancement. The first method (Sect.~\ref{sec:prop-soln:dcp}) builds a physical model of degradation and estimates the latent parameters of an underwater image from the input, and then reverses the process, like methods based on DCP. The second one (Sect.~\ref{sec:prop-soln:cyclegan}) leverages huge amount of dataset and takes advantage of deep learning methods to solve enhancement problem like image dehazing. Each method will produce different results of the enhancement and have different power consumption levels. We want to propose a new method that combines advantages of these methods by using the image quality evaluation metrics (Sect.~\ref{sec:prop-soln:metrics}). We will split methods into different categories according to the results. For algorithm described in Fig.~\ref{fig2}, we identify the quality of the underwater images and set up a metrics value as a threshold for clear images.
If it is determined as a clear image, we can send it to next step of ORB-SLAM. If not, we apply the enhancement of the first level repeatedly until the enhancement reaches that of a clear image.

\subsection{Underwater Dark Channel Prior}\label{sec:prop-soln:dcp}
As we know, the interaction between the light, the medium, and the scene will affect the quality of the underwater images. The absorption of light is related to the forward scattering and backscattering that are responsible for contrast degradation and color change in images. As the main factor is backscattering, the forward scattering can be ignored. There is a common analysis model based on underwater attenuation light modeling that uses a linear combination of the direct light and backscattering. The original Dark Channel Prior~\cite{he2011single} is based on the statistics of outdoor haze-free images. They propose that some pixels have very low intensity in at least one color channel in the RGB representation. For random image $J$, the dark channel is defined as follows,
\begin{equation}
    J^{dark}(x)=\min\limits_{y\in \Omega(x)} \min\limits_{ c \in {r,g,b}}J^c(y).
\end{equation}
Here, $J^c$ is a color channel of $J$ and $\Omega(x)$ is a local patch centered at $x$. 
There are two commutative minimum operations: one is performed on each pixel and the other is a minimum filter. The observation \(J^{dark}\to 0\) is the Dark Channel Prior~DCP. The shadows in images, colorful objects, or surface will cause a low intensity in dark channel. Also, the image model to describe the information of a haze images is shown in the following equation,
\begin{equation}
    I=J\times t +A\times (1-t),
\end{equation}
where $I$ is the observed intensity corresponding to the image captured from the camera, $J$ is the scene radiance of the clear image we need, $t$ is the medium transmission, and $A$ is the ambient light.

When applying the DCP for underwater images, it is not easy to obtain real images of underwater views in the air environment. However, the assumption that there are some pixels with very low intensity in at least one color channel also works for underwater images. For underwater images, the red channel is almost dark because of the absorption of the medium hence we can ignore the red channel information. Therefore, the underwater dark channel prior~\cite{drews2016underwater} considers the green and blue color channels. The medium transmission and the backscattering light can help restore underwater images.

\subsection{CycleGAN with SSIM Loss}\label{sec:prop-soln:cyclegan}
This method extends WaterGAN~\cite{li2018watergan,goodfellow2014generative} based on CycleGAN~\cite{zhu2017unpaired} and the SSIM loss~\cite{zhao2017loss}. It maps the color of underwater scenes to the color of air scenes without any explicit labels. The input is the underwater images, and the output is the image ``in air" that has the same content and structure. Firstly, it attempts to build a supervised model for underwater images color correction, which needs paired underwater images for the training. Secondly, it learns a cross domain mapping function between underwater images and air images. Lastly, the final model can capture context and semantic information which is the same as that of input images. To achieve this goal, the author~\cite{zhu2017unpaired} uses the unpaired image-image translation network that can use unpaired images dataset to collect special characteristics of one class and try to translate into the other class.

The input image samples are $x\in X$ (underwater) and $y\in Y$ (air). There are two adversarial discriminators $D_X$ and $D_Y$. The purpose of $D_X$ is to identify the image $x$ from the translated images $y$, while that of $D_Y$ is the opposite. The loss functions are general adversarial loss, cycle consistency loss, and SSIM loss. The adversarial loss which includes the mapping function $G:X\to Y$ and the $D_Y$, can be defined as below,
\begin{align}
L_{\text{GAN}}(G,D_{Y},X,Y)=&E_{y\sim p_{\text{data}}(y)}[\log D_{Y}(y)]\nonumber\\ &+E_{x\sim p_{\text{data}}(x)}[\log(1-D_{Y}(G(x)))].
\end{align}

Cycle consistency loss, on the other hand, aims to constrain the space of possible mapping functions, which will bring $x$ from domain $X$ back to the original images and bring $y$ from domain $Y$ back to the target image. The corresponding equation for cycle consistency loss is as below,
\begin{align} L_{\rm cyc}(G,F)=&E_{x\sim p_{\text{data}}(x)}[\Vert F(G(x))-x\Vert _{1}] \nonumber\\ &+E_{y\sim p_{\text{data}}(y)}[\Vert G(F(y))-y\Vert _{1}]. \end{align}

SSIM keeps the content and structure of the input images. The definition of SSIM between input $x$ and the translated image $G(x)$ for pixel $p$ is defined as below,
\begin{align} \text{SSIM}(p)=\frac{2\mu _{x}\mu _{y}+C_{1}}{\mu _{x}^{2}+\mu _{y}^{2}+C_{1}}\cdot \frac{2\sigma _{xy}+C_{2}}{\sigma _{x}^{2}+\sigma _{y}^{2}+C_{2}}.
\end{align}
The $p$ is the center pixel of an image patch. Then, the SSIM loss is expressed as below,
\begin{equation} L_{\text{SSIM}}(x,G(x))=1-\frac{1}{N}\Sigma _{p=1}^{N}(\text{SSIM}(p)).
\end{equation}Then the total loss is the linear combination of these three losses.

Moreover, there is another advanced method that follows a similar strategy except for the adaptive scale of the SSIM windows based on the depth map. The distances between different scenes and camera lens are different for underwater images, which will result in different levels of clarity after image restoration. To solve this problem, the authors propose a multi-scale cycle generative adversarial network~\cite{lu2019multi}. They first use DCP to get the depth information of underwater images and then apply the depth information into adversarial network training.

\subsection{Image Quality Evaluation Metrics}\label{sec:prop-soln:metrics}

Below we present three underwater image quality evaluation metrics---Underwater Color Image Quality Evaluation~(UCIQE), Structural Similarity Index~(SSIM) and Peak Signal-to-Noise Ratio~(PSNR).

\textit{UCIQE value:} 
Finding a parameter to evaluate the underwater image quality is a key step of dividing the images into different groups using different image processing techniques. Metrics corresponding to the special absorption and scattering in the underwater environment and the direct application of natural color images quality do not work. The metrics for underwater images should be related with human perception correctly, should measure different distortions of images, and be used for enhancement processing. A new UCIQE metric as a linear combination of chroma, saturation, and contrast is used in~\cite{yang2015underwater} to quantify blurring, low-contrast, and the nonuniform color cast. The UCIQE for an image is defined as,
\begin{equation} UCIQE=c_{1} \times \sigma _{c} +c_{2} \times con_{l} +c_{3} \times \mu _{s}, 
\end{equation}
\begin{figure}[t]
\centering
\includegraphics[scale=0.5]{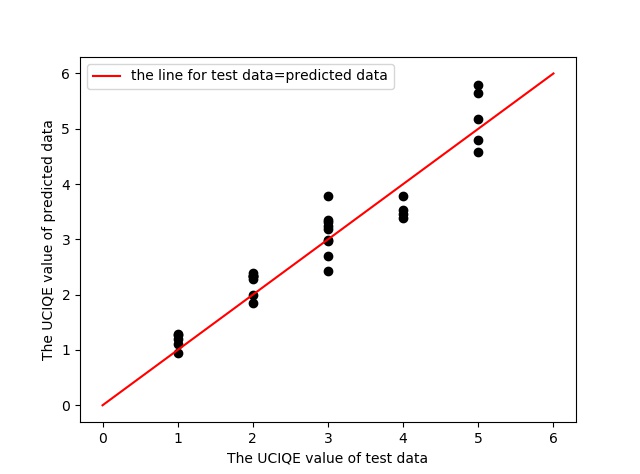}\\
\caption{\label{fig3} Model result for predicted and test data: X-axis is the value of the test data, and Y-axis is the value of the predicted data. The red line represents that the value of test data is equal to the predicted data.}
\end{figure}
where, $\sigma_{c}$ is the standard deviation of chroma, $con_{1}$ is the contrast of luminance, and $\mu_{s}$ is the average of saturation, and $c_{1},c_{2},c_{3}$ are weighted coefficients. These coefficients can be found via Multiple Linear Regression~(MLR) on training images. Firstly, 155 images are obtained from different underwater environments at different depth and turbidity levels. These images were presented to human observes who rate the images using a scale from 1 to 5---1 is the worst image (low clarity) and 5 is the best one (best clarity). To decrease the error by human subjective factor, these images will be evaluated 5 times by different human observers and the mean score will be used. After calculating the standard deviation of chroma, contrast, and the mean value of saturation, 
MLR training is performed to find the coefficients as $c_{1}=0.1654,c_{2}=0.0324,c_{3}=-0.1365$. Then, we can get a proper response variable~(UCIQE) according to three independent variables as per the linear function---$UCIQE=0.1654 \times \sigma _{c} +0.0324 \times con_{l} -0.1365 \times \mu_{s}$. In Fig.~\ref{fig3}, we show the comparison of UCIQE values of test data and predicted data (a red line is drawn to represent the best result that the predicted data is equal to the test data, for convenience). We can notice that most black points are around the red line except for few points. We can increase the size of dataset to get more accurate parameters.  

\begin{figure*}
\centering
\includegraphics[scale=0.99]{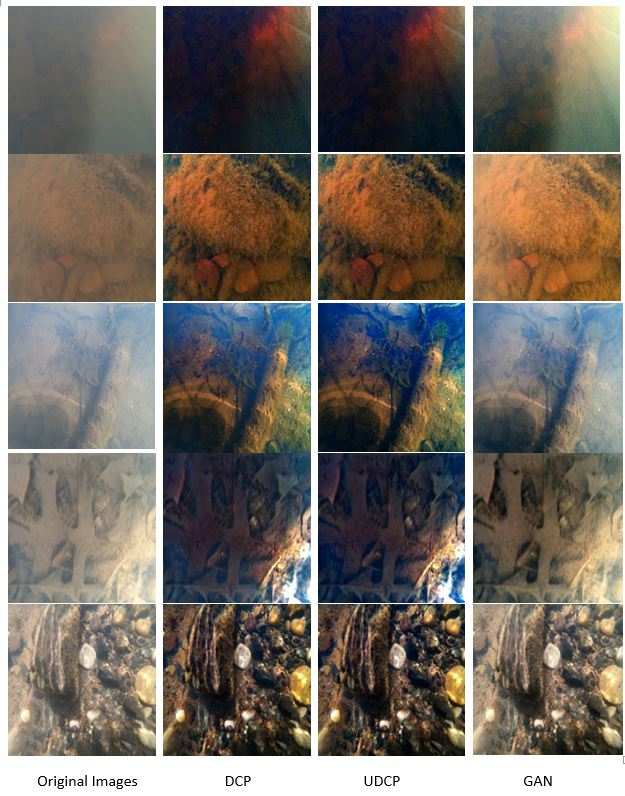}
\caption{\label{fig1}Original images (on the left) and their enhanced versions using different competing methods (DCP, UDCP, and GAN).}
\end{figure*}

\textit{SSIM metric:}
SSIM is a perception-based metric that considers image degradation in structural information. The structural information is the idea that there exists strong inter-dependencies when pixels are spatially close. Usually, these dependencies store important information about the structure of the objects. 

\textit{PSNR metric:}
PSNR is a common metric that is used to measure the quality of reconstruction of lossy compression. This ratio is defined as the maximum possible power of a signal to the noise. Logarithmic decibel scale is used to represent the value. The calculation based on the Mean Square Error~(MSE) is shown below,
\begin{equation}\label{eqn:psnr}
  \mbox{PSNR}= 10\log_{10} \frac{MAX_{I}^{2}}{MSE},
\end{equation}
\begin{equation}\label{eqmse}
  \mbox{MSE}= \frac{\sum\limits_{M,N} [I_{1}(m,n)-I_{2}(m,n)]^{2}}{M\times N}.
\end{equation}
In~\eqref{eqn:psnr}, $MAX_{I}$ is the maximum possible pixel value of the image; for the above mentioned dataset, it is $255$.
In~\eqref{eqmse}, $I_{1}$ and $I_{2}$ are $M\times N$ images, and $m,n$ are pixel indexes. 

\section{Experiment Results}\label{sec:expr}
In this section, we first present image enhancement results using our approach (Sect.~\ref{sec:expr:enh}) and then study the performance of ORB-SLAM algorithm with our enhanced images (Sect.~\ref{sec:expr:slam}).
\subsection{Image Enhancement}\label{sec:expr:enh}
\begin{figure*}[!t]
\centering
\begin{tabular}{cc}
\hspace{-5mm}  
\includegraphics[scale=0.39]{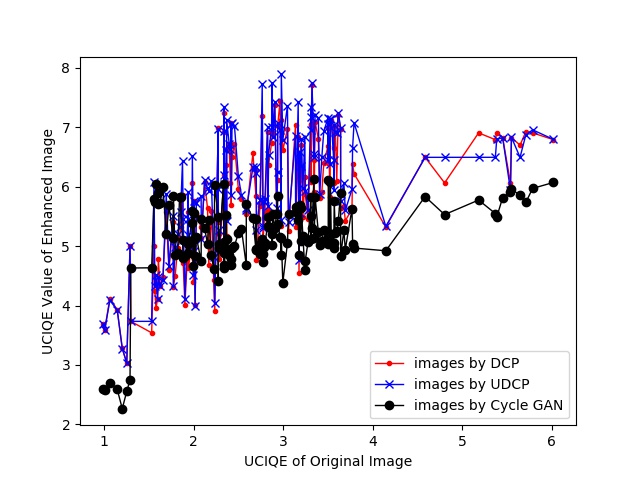}
\hspace{-0.18in}
\includegraphics[scale=0.39]{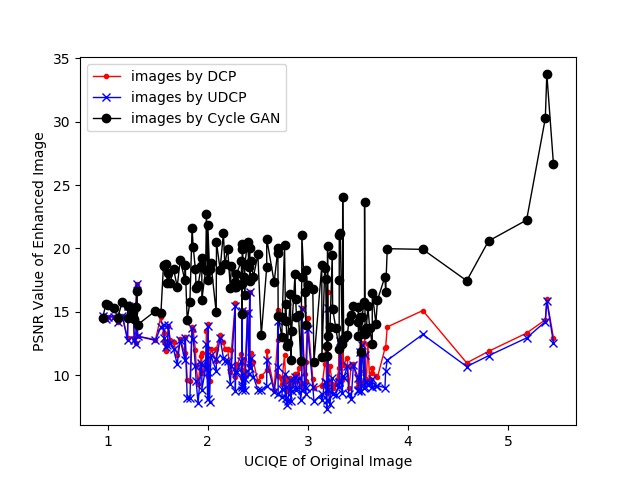}
\hspace{-0.18in}
\includegraphics[scale=0.39]{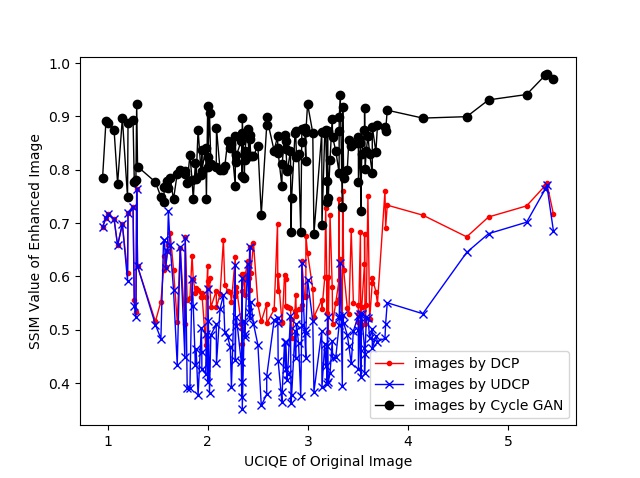}\\
\vspace{-2mm}
\hspace{-0.5cm} (a)   \hspace{5.5cm} (b) \hspace{6cm}(c)
\scriptsize
\end{tabular}
\caption{(a)~UCIQE value; (b)~PSNR value; (c)~SSIM value. We choose 150 images, using three methods to do image enhancement and calculating these three values of results. The x axis represents the UCIQE value of original images;The y axis represents UCIQE, PSNR, and SSIM value of the enhanced images.}
\label{3values}
\end{figure*}

We present experimental results corresponding to multiple rounds of experiments in the field. In the experiment---which was conducted at the Carnegie Lake and the Raritan river in New Jersey during November and December 2018---we apply the three methods to improve the quality of underwater images: the original DCP, underwater DCP, and the CycleGAN. We choose five images with different clarity levels and the results are presented in Fig.~\ref{fig1}. The first column is the original images, the second column shows the results using DCP, the third column represents the outputs using UDCP, and the last column is the result of using CycleGAN. Compared with each result, there is no method that can work very well for all the circumstances. 

For UDCP and DCP methods, they have almost similar results. In the first enhanced image, most areas are dark except for the right corner part that has excessive exposure. That is because the global dark point in original images, found by DCP and UDCP, is not suited for the whole images and the local patch is big. Therefore, There is no feature that can be captured. For UDCP method, it performs better for the water part because it only considers the blue and green color channels. At the same time, it causes more serious color distortion problem. We can see that the object's color in the enhanced images lost reality compared to the original images. For both methods, they do not work well for the first and the second images with higher turbidity so that we may need more powerful method such as GANs to improve the images. 

Among them, the CycleGAN has better image enhancement performance. When the image is very blurry such as the first image, more features can be captured when we use CycleGAN. However, the images are clear enough that using CycleGAN is too much for the whole system. For example, the last image is very clear that we can directly use in the SLAM system, although we use CycleGAN to recover it. There is no good enhancement in the results, while it adds some fake features into the images. But, this kind of situation rarely happens in rivers or lakes. Because of particles, underwater environment situations are similar to the first and second images.

Evaluating the image quality improvement cannot only depend on the human perception, and also the quality should be measured by some evaluation metrics. We select $150$ images from datasets, and calculate the PSNR, SSIM, and UCIQE values.

\begin{table}[!t]
\centering
\captionof{table}{Mean UCIQE difference in three methods.} \label{tab} 
\begin{tabular}{@{}cccc@{}}
\toprule
                & DCP    & UDCP   & Cycle GAN \\ \midrule
Mean UCIQE Difference & 2.9265 & 3.1279 & 2.3192    \\ \bottomrule
\end{tabular}
\end{table}
Fig.~\ref{3values}(a) shows the UCIQE values of original images and images enhanced by DCP, UDCP, and CycleGAN. The $x$ axis shows the values of original images, the blue line with $x$ marker represents the values of images by UDCP, the red line with dot marker stands for the values of DCP, the black line with circle marker belongs to CycleGAN. DCP and UDCP have similar values and UDCP's results are a bit better than DCP's. Images in Fig.~\ref{fig1} also show that both methods have advantages and disadvantages. For CycleGAN results, it keeps a stable values because this method transforms the image into one kind of image's style. Hence, UCIQE values will stay similar. Also, we calculate the mean difference UCIQE values between original images and enhanced images in Table~\ref{tab}. The highest increase in the performance is happened when we use UDCP method. CycleGAN may have a little overhead for images with few blur, but it shows a very good improvement for every blur image. Even UDCP shows better performance, but we choose the CycleGAN. Because the most images in test set are not under extremely environment, CycleGAN was able to handle more difficult situation.  

In Figs.~\ref{3values}(b-c), we show the PSNR and SSIM values between original images and enhanced images. For PSNR value, usually, the higher PSNR values show the better image qualities. Using CycleGAN leads to better results, although PSNR is a full-reference metric and performs poorly when it is compared with the UCIQE with human perception. For SSIM values, it performs well for in-air images, but for underwater images, it does not work better than other evaluation metrics, since there are more factors that will affect the quality of images.

\begin{figure*}[]
\centering
\begin{tabular}{cc}
\hspace{-5mm}  
\includegraphics[scale=0.47]{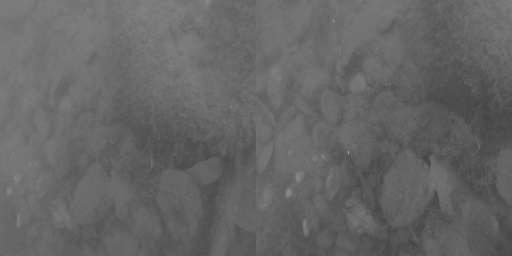}
\includegraphics[scale=0.47]{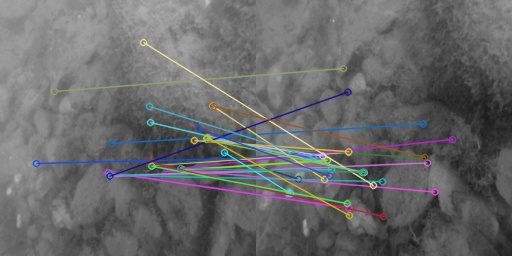}\\
\vspace{-2mm}
\hspace{-0.1cm} (a)   \hspace{8cm} (b)
\scriptsize
\end{tabular}
\caption{Result of feature matching: (a)~Before image enhancement on the original image: no feature can be matched; (b)~After image enhancement using GAN: there are 31 features matching.}\label{figfeature}
\end{figure*}

\begin{figure}[!t]
\centering
\begin{tabular}{cc}
\hspace{-2mm}  
\includegraphics[scale=0.74]{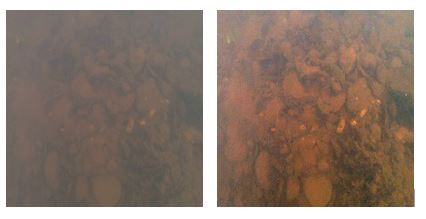}\\
\hspace{-0.6cm} (a)\\
\includegraphics[scale=0.74]{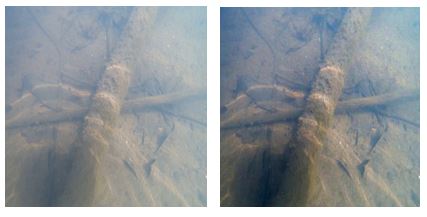}\\
\vspace{-3mm}
\hspace{-0.5cm} (b)  
\scriptsize
\end{tabular}
\caption{Image-enhancement results using CycleGAN from different locations: (a)~Bottom of the Raritan river, Somerset-New Jersey; (b)~Bottom of the Carnegie Lake, Princeton-NJ.}\label{fig6}
\end{figure}

\begin{table}[t]
\captionof{table}{The mean enhancement time for every image on different platforms}
\label{tab2} 
\begin{tabular}{l|l|l|l}
\hline
                & Linux    & Raspberry Pi & Xavier  \\ \hline
Mean\_time(sec) & 0.031761 & 0.077947    & 0.04984  \\ \hline
STD             & 0.002903 & 0.005827    & 0.006043 \\ \hline
\end{tabular}
\end{table}

We also test the algorithm for different locations, i.e., Carnegie Lake in Princeton, NJ, and the Raritan river in Somerset, NJ. Fig.~6 shows the comparisons between these two environments. As shown in the figure, the situation in river is worse than the Lake, since there are some unique stones in the lake, which are static. These kinds of features are useful for SLAM algorithms. On the contrary, for the river, there are more moving plants which will become the barriers. Therefore, the GAN can work better for river's images than lake's images.

\begin{figure}[h]
\centering
\includegraphics[scale=0.55]{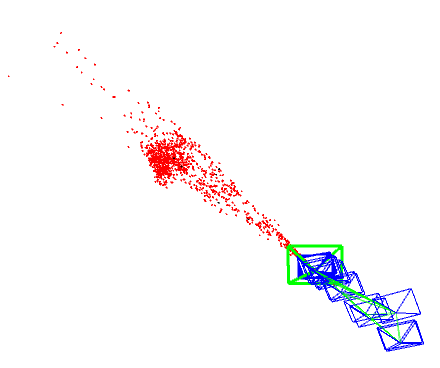}
\caption{\label{fig:orb}Graph created by ORB-SLAM.}
\end{figure}

As we know, the cost of GAN networks is another core part that we should focus on. To show the strength of this algorithms, we choose a $30~\rm{s}$ video to do the enhancement on three different platforms: Raspberry Pi 3B+, Nvidia Xavier, and a personal laptop with a Central Processing Unit~(CPU). We record the runtime of every image and calculate the mean values and standard deviation values~(std). The details are shown in Table~\ref{tab2}. We can observe the mean time is very small and the std of the values are stable and, even on Raspberry Pi, it costs $0.032$ seconds. Therefore, we can apply this algorithm to SLAM on robots which are equipped with the Raspberry Pi as the main processor.

\subsection{The ORB-SLAM Application}\label{sec:expr:slam}
The features matching is the important part of the SLAM algorithms. We compare the features matching of original images and enhanced images, as shown in Fig.~\ref{figfeature}, 
with the ORB-SLAM. We use an underwater robot to record a video from the Raritan river, NJ. Afterwards, we test this video on ORB-SLAM without image enhancement. This images from the river is blur, therefore, the SLAM cannot extract useful features to create the map and so ORB-SLAM fails to work. We use GAN to enhance the video and test the ORB-SLAM in the next step.  The result is shown in Fig.~\ref{fig:orb}, where the blue squares with $x$ mark present the key frames from the video, the green lines square present the graph, the red points show the feature and points. The robot just moves forward and the graph shows the correct direction. As it is observed in the graph, the performance of SLAM is considerably improved using the proposed solution.

\section{Conclusion and Future Work}\label{sec:conc}
Firstly, we tested multiple image-enhancement methods including a cycle generative adversarial network and dark channel prior methods. According to the experiment results, the performance of the physical model like UDCP and DCP is limited. When the image is very blurry, the results of UDCP and DCP are bellow average. However, for some light blurred images, these two methods show better results than GAN networks, while the power consumption of GAN network is huge. Furthermore, for the GAN network, the representation data is very important for the success of the model. In our proposed model, the ground truth images have similar structure and contents so that the result will have few fake features when improving images with the light blur. 

In the future, we plan to collect more images with different clarity. Moreover, given the limits we face in the underwater environment, the color of images are distorted, which is a common problem for these three methods. This leads to errors in the robot's data classification. Secondly, we have already exploited some parameters to evaluate the quality of underwater images such as UCIQE. However, the UCIQE does not always work well. Also, we will consider multiple parameters. Moreover, in the experiment section, the CycleGAN, which is a costly method, does not enhance images considerably when the images are not too blurry. We will investigate how to divide the images into different groups based on these parameters so as to decrease the cost of CycleGAN and increase the efficiency of SLAM. It is important to be able to run the algorithm on a power-limited device such as a Raspberry Pi (the main processor of many robotic platforms). 
  
\textbf{Acknowledgment:}
This work was supported by the NSF CPS Award No.~1739315.


\bibliographystyle{ACM-Reference-Format}
\bibliography{reference.bib}

\end{document}